\documentclass[preprint,12pt]{elsarticle}

\usepackage{graphicx}
\usepackage{amsmath,amssymb}
\usepackage{amsthm}

\newcommand{\dv}{\,\mathrm{div}\,}

\newcommand{\const}{\mathrm{const}}

\newcommand{\bx}{\mathbf{x}}

\newcommand{\bn}{\mathbf{n}}

\newcommand{\bJ}{\mathbf{J}}
\newcommand{\bb}{\mathbf{b}}
\newcommand{\be}{\mathbf{e}}

\newcommand{\bsigma}{\mbox{\boldmath $\sigma$}}
\newcommand{\btau}{\mbox{\boldmath $\tau$}}

\newcommand{\ba}{\mathbf{a}}
\newcommand{\bk}{\mathbf{k}}

\newcommand{\tbv}{\widetilde{\bf v}}
\newcommand{\tbB}{\widetilde{\bf B}}
\newcommand{\tba}{\widetilde{\bf a}}
\newcommand{\tbb}{\widetilde{\bf b}}

\newcommand{\bB}{\mathbf{B}}

\newcommand{\bv}{\mathbf{v}}

\newcommand{\pd}[2]{\frac{\partial #1}{\partial #2}}

\newcommand{\pdd}[3]{\frac{\partial^2 #1}{\partial #2 \partial #3}}

\journal{Physical Letters A}

\begin{document}
\graphicspath{{C:/Sergey/Papers/2009/Pictures/}}%
%{C:/Sergey/Papers/2005/SingVortInvarSubmods/Last/}
%{C:/Sergey/Papers/2005/WholeMotion/final/}{C:/Sergey/Papers/2005/Pictures/}
%{C:/Sergey/Papers/2006/Pictures/}{C:/Sergey/Papers/2006/Irrotational_Vortex/}}

\begin{frontmatter}


\ead{sergey@hydro.nsc.ru}

\title{ANALYTICAL DESCRIPTION OF STATIONARY IDEAL MHD FLOWS WITH CONSTANT TOTAL PRESSURE}


\author[a,b]{Sergey V. Golovin}

\address[a]{Lavrentyev Institute of Hydrodynamics SB RAS, 630090 Novosibirsk, Russia}
\address[b]{Department of Mechanics and Mathematics, Novosibirsk State University, 630090 Novosibirsk, Russia}

\begin{abstract}
Incompressible stationary flows of ideal plasma are observed. By introduction of curvilinear system of coordinates in which streamlines and magnetic  force lines form a family of coordinate surfaces, MHD equations are partially integrated and brought to a certain convenient form. It is demonstrated that the admissible group of Bogoyavlenskij's symmetry transformations \cite{Bogoyavl2000PhysLettA} performs as a scaling transformation for the curvilinear coordinates. Analytic description of stationary flows with constant total pressure is given. It is shown, that contact magnetic surfaces of such flows are translational surfaces, i.e. are swept out by translating one curve rigidly along another curve. Explicit examples of solutions with constant total pressure possessing a significant functional arbitrariness are given.
\end{abstract}


\end{frontmatter}


\section*{Introduction}

Ideal magnetohydrodynamics equations (MHD) describe macroscopic motions of infinitely conducting plasma under the action of internal pressure, magnetic and inertial forces. MHD approximation is applicable to wide classes of physical phenomenon all the way from plasma confinement to  astrophysical problems  \cite{Freidberg1987, Somov2006}. Well-studied in literature linear and low-dimensional reductions usually have limited areas of application \cite{Biskamp1993}. At the same time, the numerical analysis of MHD equations is complicated by the essential multi-dimensionality of MHD processes (e.g. the turbulent dynamo problem \cite{LaqndLifPit1984en}), and presence of various types of strong and weak discontinuities \cite{JeffreyTaniuti1964,Kulikovskiiatall2001}. In this connection, analytical investigations based on construction and interpretation of exact solution to ideal MHD equations play significant role. Of course, it is impossible to construct (or even to define the notion of) the general solution to MHD equations. However, particular exact solutions give opportunity to describe main features and singularities of plasma flows on the level of explicit analytical relations.

In this paper we observe the stationary ideal MHD equilibria equations:
\begin{equation}\label{SBMHD:MHD1}
\begin{array}{l}
(\bv\cdot\nabla)\,\bv-(\bB\cdot\nabla)\,\bB+\nabla P=0,\\[2mm]
(\bB\cdot\nabla)\bv=(\bv\cdot\nabla)\bB,\\[2mm]
\dv\bv=0,\;\;\;\dv\bB=0.
\end{array}
\end{equation}
Here $\bv$ is the velocity vector, $\bB$ is the magnetic field, $P=p+\frac{1}{2}|\bB|^2$ is the total pressure, $p$ is the hydrodynamical pressure of plasma. The dot denotes the scalar product in Euclidean space. Plasma density is assumed to be constant and is taken to be unity: $\rho=1$.

A significant part of known solutions to equations \eqref{SBMHD:MHD1} belong to classes of either pure ideal fluid ($\bB=0$) \cite{AKPR1998}, or static equilibrium ($\bv=0$) \cite{Freidberg1987}, or to the intermediate case of field-aligned flows ($\bv=\lambda(\bx)\bB$) \cite{GebKies1992, TassoThroum1998, ZannaChiuderi1996}, which is known to be reducible to one of two previously mentioned \cite{Bogoyavl2002}. Assuming the set of field-aligned flows to be well-analyzed, we observe the general case of plasma flows with non-collinear fields $\bv$ and $\bB$.

The induction equation (the second equation of \eqref{SBMHD:MHD1}) states, that vector fields $\bv$ and $\bB$ commute. In sections \ref{CC}, \ref{ECC} we introduce a curvilinear system of coordinates, which reduces MHD equations \eqref{SBMHD:MHD1} to a certain simpler form. A great deal of information about the flow is provided by the contact magnetic surfaces weaved out of streamlines and magnetic lines of the flow. Each of contact magnetic surfaces can be taken as an infinitely conducting impermeable wall bounding the flow. In section \ref{GI} we demonstrate, that contact magnetic surfaces are constructed in the curvilinear coordinates automatically as a part of the solution of equations.

The remarkable property of equations \eqref{SBMHD:MHD1} is that they have an infinite-dimensional symmetry group of Bogoyavlenskij's transformations \cite{Bogoyavl2000PhysLettA,Bogoyavl2002}:
\begin{equation}\label{SBMHD:Transform}
\tbv=\bB\sinh f(\varphi)+\bv\cosh f(\varphi),\quad\tbB=\bB\cosh f(\varphi)+\bv\sinh f(\varphi).
\end{equation}
Here $\varphi$ is an arbitrary function, which conserves along streamlines and magnetic force lines of the flow:
\[
(\bv\cdot\nabla)\varphi=0,\;\;\;(\bB\cdot\nabla)\varphi=0.
\]
The group allows reproduction of new exact solutions from known ones \cite{Oliveri2005}, construction of flows with current sheets and non-symmetrical equilibria states \cite{Bogoyavl2002}. It is known \cite{IlinVladimirov2004} that Bogoyavlenskij's transformations preserve the stability properties of the solution. In section \ref{ECC} we demonstrate that Bogoyavlenskij's transformations generate certain scale transformations for our curvilinear coordinates. Choice of the discontinuous scale factor allows producing a solution with a current sheet from any continuous solution of equations \eqref{SBMHD:MHD1}.

Using the reduced system of MHD equations in section \ref{CTP} we describe flows of incompressible plasma distinguished by the constant total pressure condition:  $p+\frac{1}{2}|\bB|^2=\const$. Construction of these solutions requires integration of an overdetermined system of partial differential equations, obtained by neglecting $\nabla P$ term in equations \eqref{SBMHD:MHD1}.
%In the case of pure gas dynamics ($\bB=0$) solutions with constant or time-dependent pressure were used in \cite{Zeldovich} for description of astrophysical processes of galaxies formation, and in \cite{Stanukovich} for cumulation problems.
Note, that isobaric ($p=\const$) and barochronous ($p=p(t)$) solutions to pure gas dynamics equations ($\bB=0$) are completely described in \cite{Peradzynski1990,LVO1994en,Chup1997en}. The method of integration used in these articles was based on the analysis of algebraic invariants for the Jacoby matrix $\partial\bv/\partial\bx$. In MHD equations there are two vector fields $\bv$ and $\bB$, and, correspondingly, two Jacoby matrices. The compatibility conditions of the overdetermined system are formulated in terms of algebraic invariants of both matrices and of their products in various powers. This makes the analysis cumbersome and does not allow obtaining the closed form of the solution.

In contrast to the described approach, in the curvilinear coordinates the class of solutions with constant total pressure is intrinsic and is described in section \ref{CTP} by explicit formulae with significant functional arbitrariness. We prove, that every contact magnetic surface of stationary incompressible MHD flow with constant total pressure is a translational surface, i.e., is obtained by a parallel shift of one curve along another curve in 3D space. Both curves are subjected to some restrictions. In section \ref{ES} we demonstrate examples of exact solutions describing flows in curvilinear channels. The functional arbitrariness of the solutions allows varying of the flow picture in wide extend.

\section{Curvilinear coordinates}\label{CC} In order to simplify the form of MHD equations let us introduce the following vector fields:
\begin{equation}\label{SBMHD:1}
\ba=\bv-\bB,\quad\bb=\bv+\bB,
\end{equation}
or equivalently
\begin{equation}\label{SBMHD:2}
\bv=\frac{1}{2}(\bb+\ba),\quad\bB=\frac{1}{2}(\bb-\ba).
\end{equation}
Equations \eqref{SBMHD:MHD1} give
\begin{eqnarray}\label{SBMHD:3}
&&(\ba\cdot\nabla)\bb+(\bb\cdot\nabla)\ba+2\nabla P=0,\\[2mm]\label{SBMHD:4}
&&(\ba\cdot\nabla)\bb=(\bb\cdot\nabla)\ba,\\[2mm]\label{SBMHD:5}
&&\dv\ba=0,\;\;\;\dv\bb=0.
\end{eqnarray}
Equation \eqref{SBMHD:3} by virtue of \eqref{SBMHD:4} transforms to
\begin{equation}\label{SBMHD:6}
(\ba\cdot\nabla)\bb+\nabla P=0,
\end{equation}
With new unknowns, transformation \eqref{SBMHD:Transform} takes the following form:
\[\tba=\varphi(\bx)\,\ba,\quad\tbb=\frac{1}{\varphi(\bx)}\,\bb.\]
Here $\varphi$ is the arbitrary function, which conserves along vector fields $\ba$ and $\bb$:
\[(\ba\cdot\nabla)\varphi=0,\;\;\;(\bb\cdot\nabla)\varphi=0.\]

Equation \eqref{SBMHD:4} is equivalent to the vanishing of the commutator of vector fields $\ba$ and $\bb$:
\begin{equation}\label{SBMHD:7}
[\ba,\bb]:=(\ba\cdot\nabla)\bb-(\bb\cdot\nabla)\ba=0.
\end{equation}
Under assumption of linear independence of these vector fields the commutativity condition \eqref{SBMHD:7} implies, that the fields can be taken as a coordinate basis, i.e. integral curves of the fields can serve as coordinate curves of a curvilinear coordinate system \cite{Schutz1980en}. Indeed, let us observe curvilinear coordinates $(k^1,k^2,k^3)$ such that $k^1$-- and $k^2$--coordinate curves coincide with integral curves of vector fields $\ba$ and $\bb$ correspondingly. In other words, in the space $\mathbb{R}^3$ let us define a difeomorphism
\begin{equation}\label{SBMHD:8}
\bx=\bx(\bk), \quad\det\left(\pd{\bx}{\bk}\right)\ne0
\end{equation}
such that
\begin{equation}\label{SBMHD:9}
\ba=\frac{\partial\bx}{\partial k^1},\quad\bb=\frac{\partial\bx}{\partial k^2}.
\end{equation}
Compatibility of equations \eqref{SBMHD:9} for functions $\bx(\bk)$ follows from equations \eqref{SBMHD:7}. Indeed, calculation of the mixed derivative gives
\begin{equation}\label{SBMHD:10}
\begin{array}{l}
\displaystyle\frac{\partial^2\bx}{\partial k^2\partial k^1}=\frac{\partial}{\partial k^2}\left(\frac{\partial\bx}{\partial k^1}\right)=
\frac{\partial}{\partial k^2}\bigl(\ba(\bx)\bigr)=\frac{\partial\ba}{\partial \bx}\frac{\partial\bx}{\partial k^2}=
\frac{\partial\ba}{\partial \bx}\,\bb=(\bb\cdot\nabla)\,\ba,\\[5mm]
\displaystyle\frac{\partial^2\bx}{\partial k^1\partial k^2}=\frac{\partial}{\partial k^1}\left(\frac{\partial\bx}{\partial k^2}\right)=
\frac{\partial}{\partial k^1}\bigl(\bb(\bx)\bigr)=\frac{\partial\bb}{\partial \bx}\frac{\partial\bx}{\partial k^1}=
\frac{\partial\bb}{\partial \bx}\,\ba=(\ba\cdot\nabla)\bb.
\end{array}
\end{equation}
Right-hand sides of these equations coincide by virtue of equations \eqref{SBMHD:4}, which implies the compatibility of equations \eqref{SBMHD:9}. In variables $\bx(\bk)$ the induction equation \eqref{SBMHD:4} is identically satisfied.

\section{Equations in the curvilinear coordinates}\label{ECC}
In order to rewrite equation \eqref{SBMHD:6} in the curvilinear system of coordinates it is required to calculate vector $\nabla P$ in variables $\bk$. Note, that vector $\nabla P$ in fact should be treated as a covector, transformed to vector with the use of the identity metric tensor. In other words, column-vector $\nabla P$ in equations \eqref{SBMHD:6} must be correctly written as a transposed row-vector (covector): $(\nabla P)^\ast$. Let us denote by the lower index at symbol $\nabla$ the set of variables, to which the gradient is taken. We have
\[
(\nabla_{\bx} P)^\ast=\left(\nabla_{\bk}P\,\frac{\partial\bk}{\partial\bx}\right)^\ast=
\left(\frac{\partial\bk}{\partial\bx}\right)^\ast\left(\nabla_{\bk}P\right)^\ast=
{\left(\frac{\partial\bx}{\partial\bk}\right)^{-1}}^\ast\left(\nabla_{\bk}P\right)^\ast.
\]
By multiplication of equation \eqref{SBMHD:6} from the left on the transposed Jacoby matrix and taking into account formulae \eqref{SBMHD:10} we obtain
\begin{equation}\label{SBMHD:11}
\left(\frac{\partial\bx}{\partial \bk}\right)^\ast\frac{\partial^2\bx}{\partial k^1\partial k^2}+(\nabla_{\bk}P)^\ast=0.
\end{equation}
In the coordinate representation equation \eqref{SBMHD:11} have the form
\[
\sum\limits_{j=1}^3\frac{\partial x^j}{\partial k^i}\frac{\partial^2 x^j}{\partial k^1\partial k^2}+\frac{\partial P}{\partial k^i}=0,
\quad i=1,2,3.
\]
It is remain to transform equations \eqref{SBMHD:5} to curvilinear coordinates \eqref{SBMHD:8}. To this end we use the invariance of differential 2-forms under any non-degenerate change of variables \cite{Schutz1980en}. Let us observe the following 2-forms:
\[
\begin{array}{l}
\omega^1=a^1dx^2\wedge dx^3+a^2 dx^3\wedge dx^1+a^3dx^1\wedge dx^2,\\[2mm]
\omega^2=b^1dx^2\wedge dx^3+b^2 dx^3\wedge dx^1+b^3dx^1\wedge dx^2.
\end{array}
\]
Here $a^i$ and $b^i$ are coordinates of vectors $\ba$ and $\bb$ in the Cartesian frame of reference. Equations \eqref{SBMHD:5} are equivalent to
\begin{equation}\label{SBMHD:12}
d\omega^1=0,\quad d\omega^2=0.
\end{equation}
By rewriting 2-forms $\omega^1$ and $\omega^2$ in variables \eqref{SBMHD:8}, \eqref{SBMHD:9} we obtain
\[
\omega^1=\varrho\,dk^2\wedge dk^3,\quad\omega^2=\varrho\,dk^3\wedge dk^1,\quad\varrho=\det \left(\frac{\partial \bx}{\partial \bk}\right).
\]
Closedness conditions \eqref{SBMHD:12} are equivalent to the following equations:
\begin{equation}\label{SBMHD:13}
\pd{\varrho}{k^1}=0,\quad\pd{\varrho}{k^2}=0.
\end{equation}
Thus, the original system of equations \eqref{SBMHD:MHD1} is reduced to equations \eqref{SBMHD:11}, \eqref{SBMHD:13} for unknowns $\bx=\bx(\bk)$, $P=P(\bk)$, and to representations \eqref{SBMHD:2}, \eqref{SBMHD:9} for sought vector fields $\bv$ and $\bB$. Equations  \eqref{SBMHD:13} are integrated as $\varrho=f(k^3)$ with arbitrary function $f$. Note, that by virtue of invariance of equations \eqref{SBMHD:11}, \eqref{SBMHD:13} under the change of variable $k^3\to g(k^3)$, by the suitable choice of function $g$ one can always make $f(k^3)=1$. In the final form the resulting system of equations reads
\begin{subequations}\label{SBMHD:Main}
\begin{eqnarray}\label{SBMHD:Main1}
&&\pd{\bx}{k^1}\cdot\pdd{\bx}{k^1}{k^2}+\pd{P}{k^1}=0,\\[2mm]\label{SBMHD:Main2}
&&\pd{\bx}{k^2}\cdot\pdd{\bx}{k^1}{k^2}+\pd{P}{k^2}=0,\\[2mm]\label{SBMHD:Main3}
&&\pd{\bx}{k^3}\cdot\pdd{\bx}{k^1}{k^2}+\pd{P}{k^3}=0,\\[2mm]\label{SBMHD:MainDet}
&&\det \left(\frac{\partial \bx}{\partial \bk}\right)=1.
\end{eqnarray}
\end{subequations}
System \eqref{SBMHD:Main} contains less number of equations and less number of unknowns than the original MHD equations \eqref{SBMHD:MHD1}.

Note, that system of equations \eqref{SBMHD:Main} is invariant under the change of parameters $\bk$ of the following form:
\begin{equation}\label{SBMHD:BT}
\tilde{k}^1=\varphi(k^3) k^1,\quad\tilde{k}^2=\frac{k^2}{\varphi(k^3)},\quad \tilde{k}^3=k^3
\end{equation}
with arbitrary function $\varphi$. This transformation follows from Bogoyavlenskij's transformation \eqref{SBMHD:Transform}, admitted by the original system of equations \eqref{SBMHD:MHD1}. From the point of view of introduced parametrization, formulae \eqref{SBMHD:BT} specify a scaling transformation for parameters $k^1$ and $k^2$. Besides, system \eqref{SBMHD:Main} is invariant under transformations
\begin{equation}\label{SBMHD:IT}
\tilde{k}^1=k^1+\psi(k^3), \quad\tilde{k}^2=k^2+\chi(k^3)
\end{equation}
with arbitrary functions $\psi$ and $\chi$. In what follows we threat transformations \eqref{SBMHD:BT}, \eqref{SBMHD:IT} as equivalence transformations on the set of solutions of equations \eqref{SBMHD:Main}.

\section{Geometrical interpretation}\label{GI}
Each dependence $\bx=\bx(\bk)$, $P=P(\bk)$ satisfying equations \eqref{SBMHD:Main} determines some incompressible flow of infinitely conducting plasma. A streamline of the flow $\bx=\bx_{sl}$ and a magnetic force line $\bx=\bx_{ml}$, passing through point $\bx(\bk_0)$ are parametrically defined as
\begin{multline}\label{SBMHD:SLMSL}
\bx_{sl}=\bx\left(k^1_0+\frac{1}{2}s,k^2_0+\frac{1}{2}s,k^3_0\right),\quad\bx_{ml}=\bx\left(k^1_0-\frac{1}{2}s,k^2_0+\frac{1}{2}s,k^3_0\right),\\
s\in[0,S]\subset\mathbb{R}
\end{multline}
Coordinate surfaces $k^3=\const$ are weaved out of streamlines and magnetic lines of the flow, hence, they can be treated as contact magnetic surfaces. Each of these surfaces can be taken as rigid infinitely conducting boundary, limiting the domain of the flow.

Note, that coordinate surfaces $k^3=\const$ provide an explicit form of the fibration of Euclidian space $\mathbb{R}^3$ by a family of surfaces $\psi(\bx)=\const$ introduced in section 2 of paper \cite{Bogoyavl2000PhysLettA}. This also allows us to give an explicit description of flows with contact sheets presented in \cite{Bogoyavl2000PhysLettA,Bogoyavl2002}. Indeed, Bogoyavlenskij's transformation \eqref{SBMHD:BT} does not change contact magnetic surfaces, although it transforms vectors $\ba$ and $\bb$ according to relations \eqref{SBMHD:9}, \eqref{SBMHD:BT}. The total pressure $P$ is the invariant of the transformation. Thus, by acting with the transformation \eqref{SBMHD:BT} on any continuous solution of equations \eqref{SBMHD:Main} and taking function $\varphi(k^3)$ with jump discontinuity at some value $k^3=c$, we obtain a new solution, which contains a contact discontinuity at the surface $\bx=\bx(k^1,k^2,c)$. The surface of contact discontinuity carries the electric current with the density \cite{KulikovskiLyubimov1965en}
\[\bJ=\bn\times(\bB^{(2)}-\bB^{(1)}).\]
Here $\bn$ the unit normal vector to the discontinuity surface and directed towards side ``2''; $\bB^{(i)}$ are limiting values of magnetic field vector $\bB$ on the discontinuity surface approached from side $i$. With the use of formulae \eqref{SBMHD:9} and transformation \eqref{SBMHD:BT} we obtain the following formula for the surface current:
\begin{equation}\label{SBMHD:CS}
\bJ=\frac{\varphi^{(1)}-\varphi^{(2)}}{2}\,\bn\times\left(\frac{1}{\varphi^{(1)}\varphi^{(2)}}\bx_2+\bx_1\right),\quad \bn=\frac{1}{|\bx_1||\bx_2|}\,\bx_1\times\bx_2.
\end{equation}
Hereafter, the lower index $i$ denotes the partial derivative with respect to $k^i$; sign $\times$ stands for the vector product in $\mathbb{R}^3$.  All vectors in the equation \eqref{SBMHD:CS} are calculated at $k^3=c$; parameters $k^1$ and $k^2$ are free. By $\varphi^{(i)}$ we denote the limiting values of function $\varphi(k^3)$ at $k^3\to c\pm0$.

\section{Solution with constant total pressure}\label{CTP}
Let us suppose, that total pressure $P$ is constant in all space, occupied by the flow:
\[P=\const.\]
In this case equations \eqref{SBMHD:Main1}--\eqref{SBMHD:Main3} are equivalent to
\[\pdd{\bx}{k^1}{k^2}=0\]
which implies the following representation of vector $\bx$:
\begin{equation}\label{SBMHD:14}
\bx=\bsigma(k^1,k^3)+\btau(k^2,k^3).
\end{equation}
Substitution of the representation \eqref{SBMHD:14} into equation \eqref{SBMHD:MainDet} gives
\begin{equation}\label{SBMHD:15}
\bsigma_1\cdot\bigl(\btau_2\times(\bsigma_3+\btau_3)\bigr)=1.
\end{equation}
Thus, the description of the solutions with constant total pressure is reduced to the separation of variables $k^1$, $k^2$ in equation \eqref{SBMHD:15}.

Contact magnetic surfaces for solutions with constant total pressure are defined by the equation \eqref{SBMHD:14} at fixed value of $k^3$. By virtue of this representation, these surfaces belong to a class of translational surfaces \cite{Kagan1947en}. By definition, translational surfaces are those, which can be obtained by the parallel shift of one fixed curve in 3D space along another fixed curve. In the general form translational surfaces were analyzed in  classical works by Lie, Darboux, Wirtinger, and Poincar\'{e} (see \cite{Little1983} and citations therein). As it was first noticed by S. Lie \cite{Lie1878}, any translation surface, which could be swept out in more than one way by translating one curve rigidly along another curve is completely determined by a selection of a fourth-order algebraic curve on the plane. The cited works give classification of translational surfaces, which have several independent parametrizations of the form \eqref{SBMHD:14} with fixed value of $k^3$. Translational surfaces also arise in study of billiards in rational polygons \cite{ZKa1976en}. To any such polygon there corresponds a unique translational surface such that the billiard flow in the polygon is equivalent to a geodesic flow on the surface. As far as we know, the relation between translation surfaces and flows of infinitely conducting plasma was not yet discussed in literature.

\section{Examples of solutions}\label{ES}
The general case of separation of variables in equation \eqref{SBMHD:15} is cumbersome, therefore in this paper we limit ourselves to the case of special dependence \eqref{SBMHD:14}, where vector field $\bsigma$ does not depend on $k^3$:
\begin{equation}\label{SBMHD:14a}
\bx=\bsigma(k^1)+\btau(k^2,k^3).
\end{equation}
In what follows, we denote by $\be_1$, $\be_2$, and $\be_3$ a triple of orthonormal constant vectors in $\mathbb{R}^3$. The decomposition of vector $\btau$ in the basis of vectors $\be_i$ have the following form: $\btau=\sum_{i=1}^3\tau^i\be_i$.

The dimension of the linear space $\{\bsigma\}$, spanned by vectors $\bsigma(k^1)$ for various values of $k^1$, can be 1, 2, or 3. Below we observe each of three cases separately.

{\sf a) $\dim\{\bsigma\}=1$.} In this case vector $\bsigma$ can be represented as
\[\bsigma=\alpha(k^1)\be_1.\]
By plugging this representation into \eqref{SBMHD:15} and by using the decomposition of vector $\btau$ we obtain
\[\alpha'(k^1)\left|
\begin{array}{cc}
\tau^2_2  &\tau^3_2\\
\tau^2_3  &\tau^3_3
\end{array}\right|=1.\]
Hence, accurate to equivalence transformations \eqref{SBMHD:BT}, \eqref{SBMHD:IT} this equation implies $\alpha(k^1)=k^1$, and $\tau^2_2\tau^3_3-\tau^3_2\tau^3_3=1$. Finally, we obtain the following solution:
\begin{equation}\label{SBMHD:s1}
\bx=k^1\be_1+\btau(k^2,k^3),\quad \left|
\begin{array}{cc}
\tau^2_2  &\tau^3_2\\
\tau^2_3  &\tau^3_3
\end{array}\right|=1.
\end{equation}
The function $\tau^1(k^2,k^3)$ remains arbitrary.

{\sf b) $\dim\{\bsigma\}=2$.} We choose the following representation of vector $\bsigma$:
\[\bsigma=\alpha(k^1)\be_1+\beta(k^1)\be_2\]
with linearly independent functions $\alpha$ and $\beta$. By plugging this representation into equation \eqref{SBMHD:15} we find the following relation:
\[\alpha'(k^1)\left|
\begin{array}{cc}
\tau^2_2  &\tau^3_2\\
\tau^2_3  &\tau^3_3
\end{array}\right|+\beta'(k^1)\left|
\begin{array}{cc}
\tau^3_2  &\tau^1_2\\
\tau^3_3  &\tau^1_3
\end{array}\right|=1.\]
By virtue of linear independence of functions $\alpha$ and $\beta$ this relation is satisfied (accurate to the equivalence) only when $\alpha=k^1$, $\tau^3_2\tau^1_3-\tau^1_2\tau^3_3=0$, and $\tau^2_2\tau^3_3-\tau^3_2\tau^3_3=1$. The second equation implies $\tau^1=F(\tau^3)$ with arbitrary function $F$. Hence, we obtain the solution
\begin{multline}\label{SBMHD:s2}
\bx=\Bigl(k^1+F\bigl(\tau^3(k^2,k^3)\bigr)\Bigr)\be_1+\Bigl(\beta(k^1)+\tau^2(k^2,k^3)\Bigr)\be_2+\tau^3(k^2,k^3)\,\be_3,\\[2mm] \left|
\begin{array}{cc}
\tau^2_2  &\tau^3_2\\
\tau^2_3  &\tau^3_3
\end{array}\right|=1.
\end{multline}
Here $\beta$ and $F$ are arbitrary functions of their arguments.

{\sf c) $\dim\{\bsigma\}=3$.} We use the following representation of $\bsigma$:
\[\bsigma=\alpha(k^1)\be_1+\beta(k^1)\be_2+\gamma(k^1)\be_3\]
with linearly independent functions $\alpha$, $\beta$, and $\gamma$. From equation \eqref{SBMHD:15} we obtain
\[\alpha'(k^1)\left|
\begin{array}{cc}
\tau^2_2  &\tau^3_2\\
\tau^2_3  &\tau^3_3
\end{array}\right|+\beta'(k^1)\left|
\begin{array}{cc}
\tau^3_2  &\tau^1_2\\
\tau^3_3  &\tau^1_3
\end{array}\right|+\gamma'(k^1)\left|
\begin{array}{cc}
\tau^1_2  &\tau^2_2\\
\tau^1_3  &\tau^2_3
\end{array}\right|=1.\]
Accurate to the equivalence, this implies $\alpha=k^1$,
\[\left|
\begin{array}{cc}
\tau^2_2  &\tau^3_2\\
\tau^2_3  &\tau^3_3
\end{array}\right|=1,\quad
\left|
\begin{array}{cc}
\tau^3_2  &\tau^1_2\\
\tau^3_3  &\tau^1_3
\end{array}\right|=
\left|
\begin{array}{cc}
\tau^1_2  &\tau^2_2\\
\tau^1_3  &\tau^2_3
\end{array}\right|=0.\]
These equations are satisfied only if $\tau^1=0$. The solution is given by the following formulae
\begin{equation}\label{SBMHD:s3}
\bx=k^1\be_1+\bigl(\beta(k^1)+\tau^2(k^2,k^3)\bigr)\be_2+\bigl(\gamma(k^1)+\tau^3(k^2,k^3)\bigr)\be_3,\quad
\left|
\begin{array}{cc}
\tau^2_2  &\tau^3_2\\
\tau^2_3  &\tau^3_3
\end{array}\right|=1.
\end{equation}
Here $\beta$ and $\gamma$ are arbitrary functions.

Thus, in the case of the restricted representation \eqref{SBMHD:14a} for vector $\bsigma$, the solution $\bx=\bx(\bk)$, $P=\const$ of equations \eqref{SBMHD:Main} have either of forms \eqref{SBMHD:s1}, \eqref{SBMHD:s2}, or \eqref{SBMHD:s3}.

\section{Integration of the equation for vector $\btau$}\label{IE} In all three obtained solutions components  $\tau^2$ and $\tau^3$ of vector $\btau$ are restricted by the equation
\begin{equation}\label{SBMHD:20}
\left|
\begin{array}{cc}
\tau^2_2  &\tau^3_2\\
\tau^2_3  &\tau^3_3
\end{array}\right|=1.
\end{equation}
This equation specifies an infinite-dimensional Lie group of area-preserving diffeomorphisms $(k^2,k^3)\to(\tau^2,\tau^3)$. A particular form of the diffeomorphism can be obtained by integration of linear differential equation \eqref{SBMHD:20} with respect to function $\tau^2(k^2,k^3)$ for any arbitrary function $\tau^3=\tau^3(k^2,k^3)$ or vice versa.

Below we use another method of integration of equation \eqref{SBMHD:20} in implicit form. To this end we perform an  incomplete hodograph transformation in equation \eqref{SBMHD:20}. We choose $k^3$ and $\tau^2$ as new independent variables, and $k^2$, $\tau^3$ as new unknown functions. By the differentiation of identities
\[k^2=K\bigl(k^3,\tau^2(k^2,k^3)\bigr),\quad \tau^3(k^2,k^3)=T\bigl(k^3,\tau^2(k^2,k^3)\bigr)\]
with respect to $k^2$ and $k^3$ we find the transformation of derivatives as
\[\tau^2_2=\frac{1}{K_{\tau^2}},\quad \tau^2_3=-\frac{K_{k^3}}{K_{\tau^2}},\quad \tau^3_2=\frac{T_{\tau^2}}{K_{\tau^2}},\quad \tau^3_3=\frac{T_{k^3}K_{\tau^2}-T_{\tau^2}K_{k^3}}{K_{\tau^2}}.\]
By plugging this formulae into \eqref{SBMHD:20} and canceling out the non-zero common multiplier $K_{\tau^2}$ we obtain $T_{k^3}=K_{\tau^2}$. The latter equation is solved by introduction of the potential $\Phi(k^3,\tau^2)$:
\[T=\Phi_{\tau^2},\quad K=\Phi_{k^3}.\]
Thus, the general solution of equation \eqref{SBMHD:20} is implicitly defined by formulae
\begin{equation}\label{SBMHD:stau}
\tau^3=\pd{\Phi}{\tau^2}(k^3,\tau^2),\quad k^2=\pd{\Phi}{k^3}(k^3,\tau^2)
\end{equation}
with arbitrary smooth function $\Phi(k^3,\tau^2)$. In particular, the choice
\[\Phi=\frac{1}{2}\,\tau^2\sqrt{2k^3-(\tau^2)^2}+k^3\arctan\frac{\tau^2}{\sqrt{2k^3-(\tau^2)^2}}\]
gives $\tau^2$ and $\tau^3$ as
\begin{equation}\label{SBMHD:16}
\tau^2=\sqrt{2k^3}\sin k^2,\quad \tau^3=\sqrt{2k^3}\cos k^2.
\end{equation}
Note, that any solution $(\tau^2,\tau^3)=\bigl(\tau^2(k^2,k^3),\tau^3(k^2,k^3)\bigr)$ of equation \eqref{SBMHD:20} can be modified by either of the following ways:
\[\widetilde{\tau}^2 = \tau^2+G(\tau^3), \quad\mbox{ or }\quad \widetilde{\tau}^3 = \tau^3+G(\tau^2)\]
with arbitrary function $G$.

\section{Interpretation of obtained solutions}\label{IS}
For the translational surface specified by equation \eqref{SBMHD:14} with fixed value $k^3=c$ the curve $\bx=\bsigma(k^1,c)$ will be referred to as the generator, and the curve $\bx=\btau(k^2,c)$ as the directrix.

Let us observe solution \eqref{SBMHD:s1}. The generator here is a straight line along vector $\be_1$, hence contact magnetic surfaces $k^3=c$ are nested cylinders. The directrix of cylinders is swept by the vector $\bx=\btau(k^2,c)$ whose first component $\tau^1(k^2,k^3)$ can be chosen arbitrarily, and two remaining components should satisfy equation \eqref{SBMHD:stau}. In particular, choice of functions $\tau^2$, $\tau^3$ in accordance to \eqref{SBMHD:16} specifies usual circular cylinders.

Note, that arbitrary choice of function $\tau^1(k^2,k^3)$ does not change contact magnetic surfaces. However, the choice of function $\tau^1$ considerably modifies the picture of magnetic lines and streamlines \eqref{SBMHD:SLMSL} of the flow. In particular, by the suitable choice of function $\tau^1$ plasma flow with fixed contact magnetic surfaces can be made either sub-alfv\'{e}nic ($|\bv|<|\bB|$), alfv\'{e}nic ($|\bv|=|\bB|$), or super-alfv\'{e}nic ($|\bv|>|\bB|$).

\begin{figure}[t]
  \begin{center}
  \includegraphics[width=0.7\textwidth]{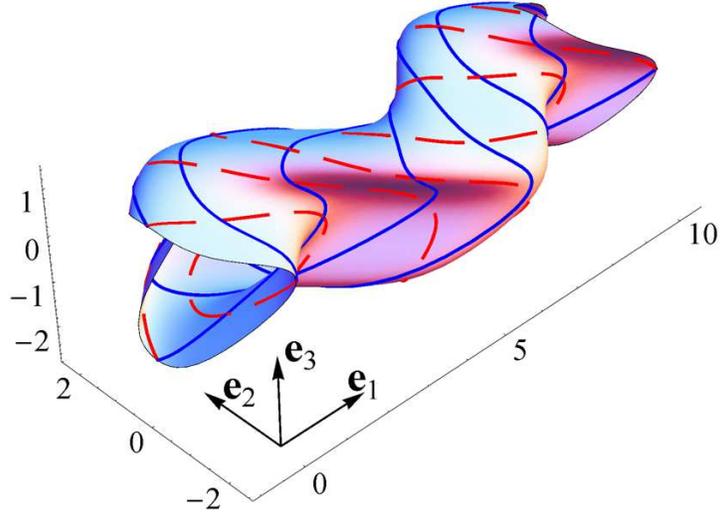}
  \caption{The contact magnetic surface defined by solution \eqref{SBMHD:s2} with $\beta=\sin k^1$, $\tau^2=\sqrt{2k^3}\sin k^2$, $\tau^3=\sqrt{2k^3}\cos k^2$, $k^3=1$, $F=\cos2\tau^3$. Continuous and dashed curves on the surface are magnetic force lines and streamlines respectively.}\label{fig1}
  \end{center}
\end{figure}

In the solution \eqref{SBMHD:s2} the generator is an arbitrary planar curve specified by function $\beta(k^1)$. The directrix belongs to a plane spanned by vectors $\be_2$ and $\be_3$, and is defined by a solution of equation \eqref{SBMHD:20} . This curve can be stretched along $\be_1$-direction by a choice of a non-zero function $F(\tau^3)$. Example of the contact magnetic surface is given in figure \ref{fig1}.

Finally, in the solution \eqref{SBMHD:s3} the generator is a 3D curve specified by functions $\beta$ and $\gamma$. The directrix is a planar curve given by the solution of equation \eqref{SBMHD:20}. For this solution the modification of a flow is possible due to the arbitrariness in the choice of the generator and in the solution of equation \eqref{SBMHD:20} for the directrix.

\section{Summary} For incompressible stationary flows of ideal plasma we introduce a convenient curvilinear coordinate system \eqref{SBMHD:8}, \eqref{SBMHD:9}, which allows partial integration of MHD equations in the form \eqref{SBMHD:Main} and provides a natural geometrical description of the flow. For any solution of equations  \eqref{SBMHD:Main} we give explicit formulae of streamlines and magnetic force lines \eqref{SBMHD:SLMSL}. We show, that Bogoyavlenskij's infinite-dimensional group of symmetry transformations reduces to a group of scale transformations \eqref{SBMHD:BT} for the curvilinear coordinates. The group preserves contact magnetic surfaces of the flow. Bogoyavlenskij's transformation with discontinuous scale multiplier produces flow with current sheet from any continuous flow of plasma. The surface current is explicitly given by formula \eqref{SBMHD:CS}.

By using the curvilinear coordinates we describe stationary flows of ideal incompressible plasma with constant total pressure. It is shown, that contact magnetic surfaces of such flows are translational surfaces, i.e. are swept out by translating one curve rigidly along another curve. We give  explicit examples \eqref{SBMHD:s1}--\eqref{SBMHD:s3} of solutions with constant total pressure. The significant functional arbitrariness of solutions allows significant modification of described flows.

\section*{Acknowledgements.} The work was partially supported by RFBR (grant no. 08-01-00047), by Ministry of Education and Science of Russian Federation (project no. 2.1.1/3543) and by the Russian Academy of Sciences (project 2.14.1).



%\bibliographystyle{amsplain}
%\bibliography{Reference}
%
%\end{document}

\end{document}